\documentclass[10pt,conference]{IEEEtran}
\IEEEoverridecommandlockouts
\usepackage{cite}
\usepackage{color}
\usepackage{amsmath,amssymb,amsfonts}
\usepackage{algorithmic}
\usepackage{graphicx}
\usepackage{array}
\usepackage{tabularx}
\usepackage{url}
\usepackage{textcomp}
\usepackage{xcolor}
\usepackage{subfigure}
\usepackage{bm}
\usepackage{algorithm}
\def\BibTeX{{\rm B\kern-.05em{\sc i\kern-.025em b}\kern-.08em
		T\kern-.1667em\lower.7ex\hbox{E}\kern-.125emX}}
\begin{document}
	
	\title{Deadline Aware Two-Timescale Resource Allocation for VR Video Streaming}
	
	\author{\IEEEauthorblockN{Qingxuan~Feng\IEEEauthorrefmark{1},~Peng~Yang\IEEEauthorrefmark{1},~Zhixuan~Huang\IEEEauthorrefmark{1},~Jiayin~Chen\IEEEauthorrefmark{2},~and~Ning~Zhang\IEEEauthorrefmark{3}}
		\IEEEauthorblockA{\IEEEauthorrefmark{1}School of Electronic Information and Communications, Huazhong University of Science and Technology, Wuhan, China \\
			\IEEEauthorrefmark{2}Department of Electrical and Computer Engineering, University of British Columbia Vancouver, Canada\\	
			\IEEEauthorrefmark{3}Department of Electrical and Computer Engineering, University of Windsor, Windsor, ON, Canada\\
			Email: \IEEEauthorrefmark{1}\{qxfeng, yangpeng, zhixuan\_huang\}@hust.edu.cn,
			\IEEEauthorrefmark{2}jiayin.chen2018@gmail.com,
			\IEEEauthorrefmark{3}ning.zhang@uwindsor.ca}
	}
	\maketitle
	
	\begin{abstract}
		In this paper, we investigate resource allocation problem in the context of multiple virtual reality (VR) video flows sharing a certain link, considering specific deadline of each video frame and the impact of different frames on video quality. Firstly, we establish a queuing delay bound estimation model, enabling link node to proactively discard frames that will exceed the deadline. Secondly, we model the importance of different frames based on viewport feature of VR video and encoding method. Accordingly, the frames of each flow are sorted. Then we formulate a problem of minimizing long-term quality loss caused by frame dropping subject to per-flow quality guarantee and bandwidth constraints. Since the frequency of frame dropping and network fluctuation are not on the same time scale, we propose a two-timescale resource allocation scheme. On the long timescale, a queuing theory based resource allocation method is proposed to satisfy quality requirement, utilizing frame queuing delay bound to obtain minimum resource demand for each flow. On the short timescale, in order to quickly fine-tune allocation results to cope with the unstable network state, we propose a low-complexity heuristic algorithm, scheduling available resources based on the importance of frames in each flow. Extensive experimental results demonstrate that the proposed scheme can efficiently improve quality and fairness of VR video flows under various network conditions.
	\end{abstract}
	
	\section{Introduction}
	The Metaverse is anticipated to be the next generation of the Internet, allowing individuals to entertain and socialize in a virtual space. Virtual reality (VR) is a crucial technology for implementing the Metaverse and has attracted considerable attention\cite{9877927}. It is predicted that VR revenues are expected to climb to 87.97 billion USD in 2025\cite{9388911}, demonstrating the great promise of VR market. As a key component in VR applications, VR video can provide users with immersive experience due to the panoramic feature of 360-degree visual information. Users can freely move their heads to explore video content in different directions through the VR helmet.
	
	Due to the near-eye display on the helmet, VR videos are required to be encoded at high resolution to provide immersive viewing experience, resulting in prohibitively high data volume. Generally, a $6$K VR video can consume more than $4\times$ higher bandwidth than 2D video with the same quality of experience (QoE)\cite{9514569}. Besides, delayed video content, low resolution or frame rate all cause VR sickness\cite{yangss,9416950}. Thus, VR video demands high on low end-to-end (E2E) delay and bandwidth resources. Considering that FoV size is $90^\circ \times 90^\circ$ according to human eye feature system, some tile-based adaptive VR video streaming schemes have been proposed to improve user's QoE\cite{9838819,9351629,huang}, which adjust the video bitrate depending on the estimation of bandwidth and FoV. However, the maximum bandwidth that a VR video flow can achieve is affected by other flows sharing the same link. When the bandwidth resource on a certain link is insufficient, referred to as the bottleneck link, adaptive transmission schemes are designed to reduce video bitrate to avoid stalling, leading to degraded QoE\cite{aljoby2021diffperf}.
	
	With the high bandwidth requirement and popularity of VR video, the probability of multiple VR video streaming sharing bottleneck link increases. In this context, the network congestion will occur, resulting in increased E2E delay and wasted bandwidth resources due to transmitting expired data (e.g., a tile predicted to be in the FoV turns out to be not viewed by the user due to head movement stochastics). To address the issue, some resource scheduling schemes have been proposed to improve resource utility\cite{9388911,9488771,9741351}, which regard the last-mile transmission link as bottleneck due to the popularity of mobile edge computing\cite{8466606,9380662}. However, with the development of 5G, mmWave communication is applied to wireless VR transmission due to the ultra-high data rate, alleviating the problem of limited bandwidth resources on wireless link, while wired link may turn out to be the bottleneck\cite{chakareski2023millimeter,yan2019transmission}. Existing resource scheduling algorithms for wired link are employed on the switch and mainly to meet quality-of-service (QoS) requirements of different flows\cite{9667095,9187995}, where the traffic characteristics are taken into account to allocate resources. Although these algorithms can effectively reduce E2E delay, it is not suitable to apply them in VR video transmission due to the following reasons. Firstly, these methods ignore the spatial and temporal differential of packets caused by video encoding and viewpoint position, while the impact of different VR video packets on user's QoE is variable. Secondly, fixed delay requirement conflicts with the nature of video service where each frame has a distinct deadline\cite{yang2021predictive}. Especially, the methods may perform poor when congestion is serious because the packets violating the delay requirement are forwarded, leading to unnecessary bandwidth waste.
	
	In this work, we aim at enhancing the quality of multi-user VR video streaming, and propose a deadline-aware resource allocation scheme. Firstly, we consider specific deadline of each video frame and estimate the queuing delay bound via historical information to drop expired packets in advance. Meanwhile, to capture the impact from differential of VR video packet, we jointly consider user viewport characteristics and encoding method to model the importance of each frame, based on which we sort arriving frames for each flow. Then we formulate an optimization problem to minimize long-term quality loss caused by frame dropping. Due to the different granularity of frame dropping frequency and network condition fluctuation, we propose a two-timescale resource allocation scheme, including minimum quality assurance resource allocation in long timescale and fine-grained resource scheduling in short timescale. The main contributions of this paper are summarized as follows:
	\begin{itemize}
		\item We establish a queuing delay bound estimation model. Through adding video frame information into packet header, the bottleneck node can calculate current queuing delay bound and save bandwidth resources by dropping invalid packets proactively.
		\item We propose a frame importance model based on user's viewpoint and encoding method. Based on the frame importance and queuing delay bound, we design a sort method to ensure that frame with higher importance and stricter delay bound can be transmitted earlier.
		\item We propose a two-timescale resource allocation scheme to minimize long-term quality loss. The long timescale resource allocation exploits queuing theory to guarantee minimum quality requirements. Furthermore, to cope with varying network and improve resource utilization, a heuristic algorithm is adopted in a short timescale.
	\end{itemize}
	
	The remainder of the paper is organized as follows. Section II describes system model and problem formulation. The two-timescale resource allocation scheme is detailed in Section III. Then experimental results are shown in Section IV. Finally, we conclude this work in Section V.
	\section{System Model and Problem Formulation}
	\subsection{System Overview}
	We focus on a single-bottleneck node scenario, where multiple VR video flows share bandwidth resource of the bottleneck node. In this scenario, we assume the storage resources are sufficient to queue the packets. Thus, the VR video flows will not turn to other paths during the transmission process. The set of flows is denoted by $\mathcal F$. Based on common VR video transmission framework, each flow is composed of temporally continuous chunks $\mathcal C=\left\{1,2,\cdots,C\right\}$ and each chunk contains multiple independent tiles $\mathcal M=\left\{1,2,\cdots,M\right\}$. Generally, a tile consists of a group of pictures (GoP), denoted by $\mathcal K=\left\{1,2,\cdots,K\right\}$. For simplicity, we use frame $(c,m,k)$ to describe frame $k$ of tile $m$, chunk $c$. Because a frame may have multiple packets and the loss of any packet will cause the frame to fail to decode, the packets belonging to the same frame are regarded as a whole.
	
	The framework of our system is depicted in Fig. 1. In the ingress phase, the bottleneck node receives VR data from different flows and puts them into corresponding queue. Meanwhile, based on the information provided by the servers, i.e., round-trip time (RTT), deadline and importance of each frame, the bottleneck node estimates queuing delay bound and sorts frames of each flow. In the egress phase, the queue scheduling module allocates resources for all flows. We consider a discrete-time system with scheduling intervals of the same duration $\delta$. At the beginning of the $n$-th interval, the queue scheduler allocates resource $B_f^n$ for flow $f$. Then a frame-level deficit weighted round robin algorithm is employed for frame forwarding module, where the updated deficit $x_f^n$ of each flow at $n$-th interval depends on the allocation results, i.e., $x_f^n =\delta B_f^n$. Besides, the frame forwarding module drops frames whose queuing time exceed the delay bound and record the queuing delay of forwarded frames. The client receives video data and regularly reports user status information to the server, i.e., user viewpoint and viewing status.
	\begin{figure}[t]
		\centering
		\setlength{\abovecaptionskip}{-0.1cm}
		\includegraphics[width=\linewidth]{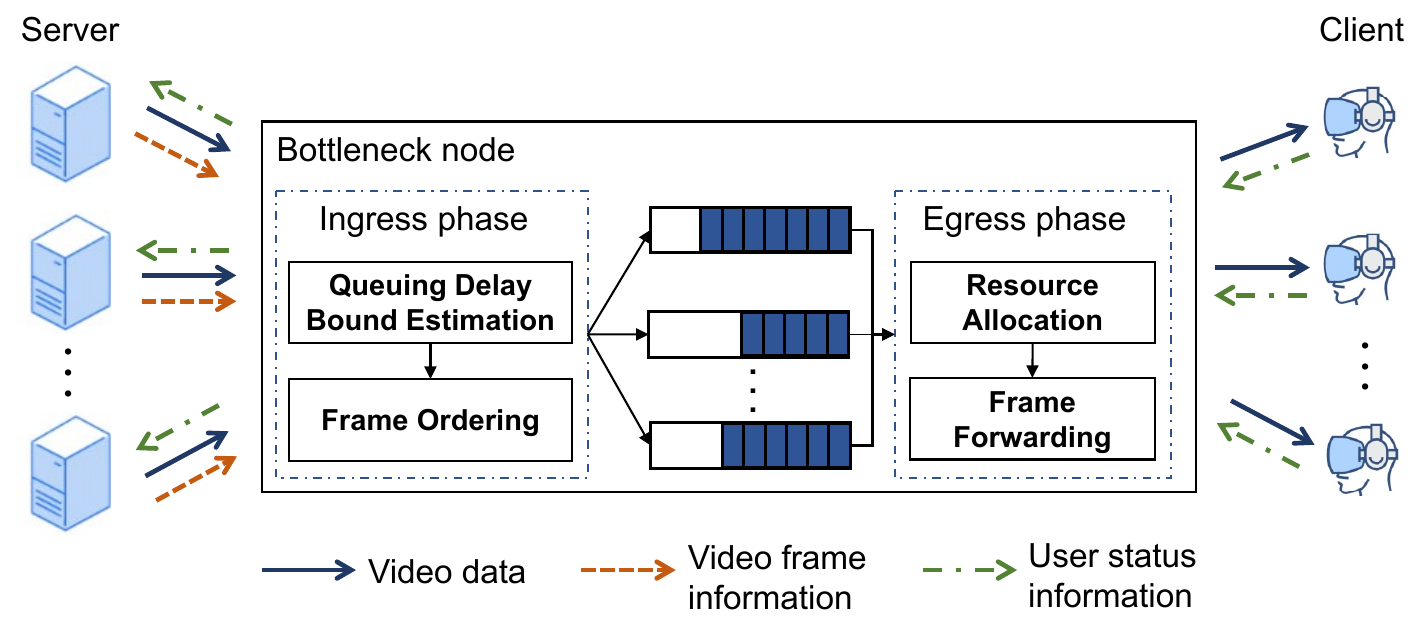}
		\caption{An illustration of the system model. }
		\label{A framework of the system }
		\vspace{-0.5cm}
	\end{figure}
	\subsection{Queuing Delay Bound Model}
	We consider the pull-based VR video flow. The client sents a request for each chunk to the server, along with current viewing status. Suppose a request for the $c$-th chunk is sent when the user is watching the $\hat{c}$-th chunk, then the deadline of chunk $c$ should be the playback interval between the two chunks, i.e., $DDL_{c} = a(c-\hat{c})$, where $a$ is chunk duration. To obtain the queuing delay bound, we need to measure the delay of three links, i.e., transmission delay of the request, delay from the server to the bottleneck node and delay from the bottleneck node to the client. Assuming that the network condition is stable in the short term, the value of historical RTT and queuing delay can be adopted to estimate the total delay of the three links, i.e., $\widetilde{RTT}-\widetilde{q}$, where $\widetilde{q}$ and $\widetilde{RTT}$ mean exponentially weighted moving average (EWMA) value of $q$ and $RTT$. Note that the RTT and queuing delay must be acquired from the same set of packets. 
	
	To satisfy the requirement, the server should mark frame deadline, frame index and RTT value based on acknowledge information of all of the packets of the latest frame. The marks are added into the option field of TCP packet header\cite{eddy2022rfc}, namely a 4-byte frame index, a 2-byte frame deadline, a 2-byte RTT value and a 1-byte RTT index indicating which frame used to calculed RTT, denoted by $(c',m',k')$. The deadline of frame $(c,m,k)$ is $DDL_{c,m,k} = DDL_{c}-(\mathsf{t}^{(s)}_{c,m,k}-\mathsf{t}^{(s)}_{c,1,1})$, where $\mathsf{t}^{(s)}_{c,m,k}$ is the time when frame $(c,m,k)$ is sent. Besides, to distinguish VR video packets and other types of packets, e.g., the request sent by client, a 1-bit flag is added into the option field. Then, the bottleneck node can estimate the queuing delay bound of frame $(c,m,k)$ by packet header information and recorded historical queuing delay, that is, 
	\begin{equation}
		D_{c,m,k}=DDL_{c,m,k}-(\widetilde{RTT}_{c',m',k'}-\widetilde{q}_{c',m',k'}).
	\end{equation}
	\subsection{Frame Importance Model}
	We consider frames with different importance based on encoding mode and user's FoV. In widely used H.265 encoding method\cite{sullivan2012overview}, there are three types of frames, i.e., I-frame, P-frame and B-frame. A GoP consists an I-frame, multiple P-frames and B-frames. I-frame adopts intra frame coding and thus is decoded independently. P-frame and B-frame are encoded with reference to the other frames. Thus, frames in a GoP are divided into multiple temporal sub-layers based on the reference relationship. Coding low sub-layer frames does not require the content of high sub-layer frame. Contrarily, coding high sub-layer frames require the content of low sub-layer frame. Fig. 2 is an example of two temporal sub-layers, where the index means the encoding order. We define $N_k$ as the number of frames that cannot be decoded if any packet of frame $k$ is dropped. For instance, $N_1$ is $N_{GoP}-1$ in Fig. 2, where $N_{GoP}$ is the number of frames in a GoP. Obviously, the frame with a larger $N_k$ is more important to user's QoE.
	
	Besides decoding dependency, user's viewpoint position also has significant influence on QoE. According to human visual characteristics, user is more sensitive to quality in the viewpoint region, and less sensitive in the peripheral region. Thus, we define the importance of frame $(c,m,k)$ as the product of viewing probability and decoding dependency proportion, denoted by $\gamma_{c,m,k}$, which is written as
	\begin{equation}
		\gamma_{c,m,k}=p_{c,m}\frac{N_k}{N_{GoP}},
	\end{equation}
	where $p_{c,m}\in \left(0,1\right]$ is the viewing probability of tile $(c,m)$, acquired by viewpoint prediction in the server.
	\begin{figure}[t]
		\centering
		\setlength{\abovecaptionskip}{0.cm}
		\includegraphics[width=0.65\linewidth]{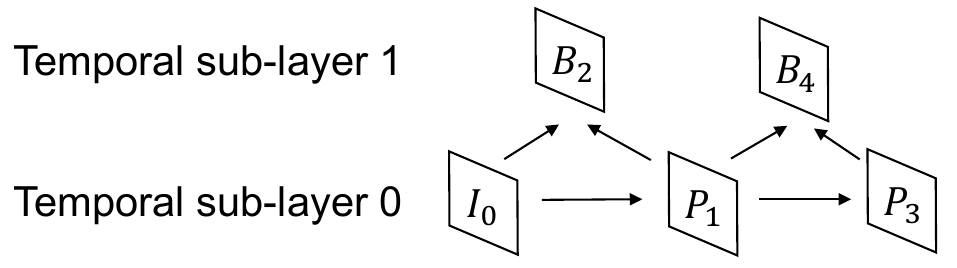}
		\caption{An example of encoding order of consecutive video frames. }
		\label{an example of encoding ordering }
		\vspace{-0.5cm}	
	\end{figure}
	\subsection{Problem Formulation}
	We consider the quality loss caused by the frames that are dropped due to queuing delay bound violation. Denote $\mathcal {Q}_f^n$ by the set of ordered frames of flow $f$ at the $n$-th interval. For simplicity, we define frame $\mathsf{F}_i$ as the $i$-th frame in set $\mathcal {Q}_f^n$. Denote $O_i^{n}$ by the tolerable queuing time of frame $\mathsf{F}_i$ at the end of $n$-th interval, which can be calculated as
	\begin{equation}
		O_{i}^{n} = D_{i}-(n\delta-\mathsf{t}_{i}^{(a)}),
	\end{equation}
	where $D_i$ is the queuing delay bound of frame $\mathsf{F}_i$ and $n\delta-\mathsf{t}_{i}^{(a)}$ represents the elapsed time from the arrival of the first packet of frame $\mathsf{F}_i$. Because the frame with lower tolerable queuing time and higher importance in the same flow should have higher forwarding priority, we define the weight of frame as a linear combination of tolerable queuing time and importance, i.e, $w_{i}=\gamma_{i}-\beta O_{i}^{n-1}$, where $\beta$ is the coefficient of tolerable queuing time. And the frames in set $\mathcal {Q}_f^n$ are arranged in descending order of weight.
	
	When $\mathcal {Q}_f^{n}$ is determined, whether a frame can be forwarded depends on the allocated resource and the tolerable queuing time. Then the forwarded frames set of flow $f$ at the $n$-th interval can be expressed by
	\begin{equation}
		\mathcal {A}_f^{n}=\left\{\mathsf{F}_j\in \mathcal {Q}_f^{n}:\sum\limits_{i=1}^{j}s_{i}\mathbb{I}(O_{i}^{n})\leq \delta B_f^n, O_{j}^{n}\geq 0\right\},
	\end{equation}
	where $s_i$ is the data size and $\mathbb{I}(\cdot)$ is an indicator function. When $O_{i}^{n}\geq 0$, $\mathbb{I}(O_{i}^{n})=1$, which means queuing delay bound is satisfied, otherwise $\mathbb{I}(O_{i}^{n})=0$. Correspondingly, the dropped frames set of flow $f$ at the $n$-th interval can be expressed by
	\begin{equation}
		\overline{\mathcal {A}}_f^{n}=\left\{\mathsf{F}_i\in \mathcal {Q}_f^{n} \backslash\mathcal {A}_f^{n}:O_{i}^{n}<0 \right\},
	\end{equation}
	Based on the frame importance model, we define the quality loss function at each interval as the ratio of the importance of dropped frame to that of departing frames, i.e., $L_f^n = \sum_{\mathsf{F}\in \overline{\mathcal {A}}_f^n}\gamma_{\mathsf{F}}/\sum_{\mathsf{F}\in \mathcal {A}_f^n \cup \overline{\mathcal {A}}_f^n} \gamma_{\mathsf{F}}$.
	Finally, our objective is to minimize quality loss of all flows in a long term, that is,
	\begin{equation}
		\begin{aligned}
			\min\limits_{\left\{B_f^n\right\}_{f\in \mathcal F}} & \sum_{n=1}^{N} \sum_{f\in \mathcal F} L_f^n \\
			s.t. \quad & C1: \sum\limits_{f \in \mathcal F} B_f^n \leq B \\
			& C2: L_f^n \leq \epsilon
		\end{aligned}
	\end{equation}
	where $N$ means the interval when all flows end. C1 is total bandwidth resources constraint and C2 guarantees the minimum quality requirement of each flow. Note that quality loss can only be calculated when the duration of allocation interval is not less than queuing delay bound. Otherwise, the quality loss will be zero regardless of the resource allocation results as no frames are dropped. Since the duration of a VR video chunk is usually one second, the queuing delay bounds of different frames vary on the second scale, which calls a long timescale allocation. However, network fluctuation will cause the queuing bound deviating from the true value, which occurs in the millisecond scale. Besides, to solve the problem, we need to know the queuing bounds of frames in queue. Thus, a two-timescale resource allocation scheme is proposed, where the short timescale is to cope with variable network state.
	\begin{figure}[t]
		\centering
		\setlength{\abovecaptionskip}{-0.12cm}
		\includegraphics[width=1\linewidth]{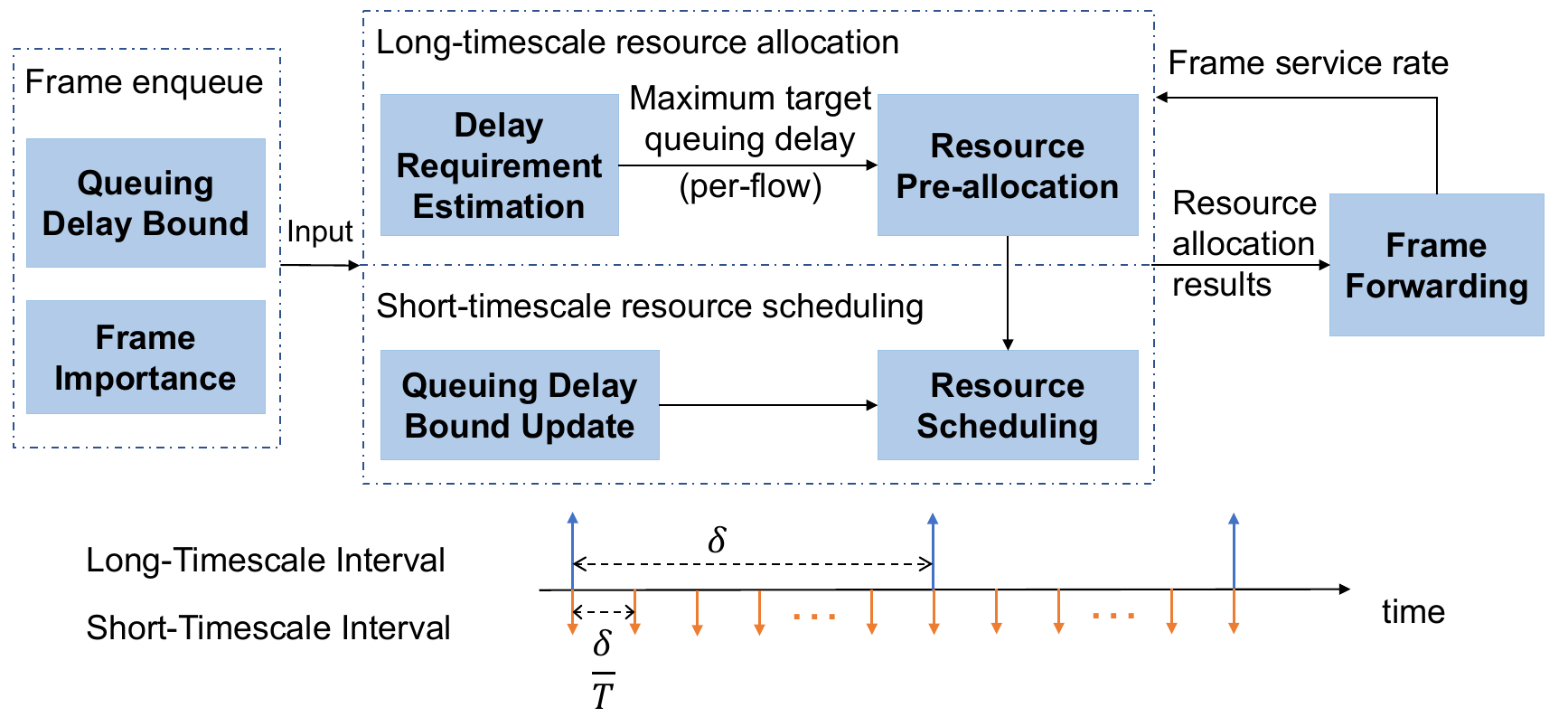}
		\caption{Illustration of two-timescale resource allocation scheme. }
		\label{Illustration of two-timescale resource allocation scheme }
		\vspace{-0.55cm}
	\end{figure}
	\section{Two-timescale resource allocation scheme}
	In this section, we present a two-timescale resource allocation scheme to improve the quality of different VR flows. As shown in Fig. 3, the resource allocation process contains two phases, i.e., resource allocation on a long timescale and resource scheduling on a short timescale. For simplicity, we denote the $n$-th long-timescale interval (LTI) and $t$-th short-timescale interval (STI) within $n$-th LTI by the superscript $n$ and $n,t$. Each LTI consists of $T$ STIs and the duration of an STI is $\delta/T$. The former estimates maximum target queuing delay for each flow and determines the allocation resources to guarantee the constraint of minimum quality requirement. The latter updates the queuing delay bounds for all frames in the queue and schedules the available resources to maximize the utilization. Frame forwarding module receives the total allocation results and continuously forwards frame.
	\subsection{Long-Timescale Resource Allocation}
	At the end of $n$-th LTI, the quality loss of all flows is calculated based on the importance of frames departing during the interval. Since the queuing delay bound will be estimated and recorded when a frame arrives, we can obtain the set of frames of flow $f$ destined to depart in this interval based on the tolerable queuing time of each frame, that is,
	\begin{equation}
		\mathcal {S}_f^{n}=\left\{\mathsf{F}_j\in \mathcal {Q}_f^{n}: O_{j}^{n-1}\leq \delta \right\}.
	\end{equation}
	To satisfy the minimum quality requirement, we can estimate the frame queuing delay $d^n_f$ that make constraint C2 hold based on the importance and delay bounds of frames in set $\mathcal {S}_f^{n}$. Specially, $d^n_f$ is the maximum value that satisfies the inequality $\sum_{\mathsf{F}\in  \mathcal {S}_f^{n}}\gamma_{\mathsf{F}}\mathbb{I}(d-D_{\mathsf{F}})/\sum_{\mathsf{F}\in \mathcal {S}_f^{n}} \gamma_{\mathsf{F}} \leq \epsilon$, where $D_{\mathsf{F}}$ is the queuing delay bound of frame $\mathsf{F}$. Although delay is a continuous variable, its estimation can be obtained by dichotomy search, where the search range is the queuing delay bounds of frames in the set $\mathcal {S}_f^{n}$. The longer queuing delay, the more frames will be dropped, leading to a higher quality loss. So we define $d_f^n$ as maximum target queuing delay of flow $f$.
	
	To obtain the resource allocation decision based on $d_f^n$, we consider frame arrival and service processes for each flow and adopt Kingman formula to control the queue. Kingman formula is widely utilized to estimate the mean queuing delay in a G/G/1 queue\cite{kingman1961single}, which is applicable to the video frame transmission owing to the frame size following Gamma distribution. For simplicity, the index of flow is omitted and the expectation of queuing delay of each flow is
	\begin{equation}
		E(d_{queue}) \approx (\frac{\rho}{1-\rho})(\frac{c_a^2+c_s^2}{2})\mu_s,
	\end{equation}
	where $c_a=\frac{\sigma_a}{\mu_a}$ and $c_s=\frac{\sigma_s}{\mu_s}$ are the coefficient of variation for arrival and service times. $(\mu_a,\sigma_a)$ and $(\mu_s,\sigma_s)$ are the mean and standard deviation of arrival and service times. $\rho=\frac{\mu_s}{\mu_a}$ is the utilization. By setting the $E(d_{queue})$ to target queuing delay $d^n_f$, we can estimate the required service rate. Finally, the allocation resource can be calculated by
	\begin{equation}
		\hat{b}^n_f=\frac{s_{ave}}{\mu_s} = s_{ave} \frac{\sqrt{1+2\mu_a(c_a^2+c_s^2)/d_f^n}+1}{2\mu_a},
	\end{equation}
	where $s_{ave}$ means the average frame size in the set $\mathcal {S}_f^{n}$. When the target queuing delay decreases, more bandwidth resources should be allocated to the flow to guarantee the quality requirement. Besides, the allocation result is also related to the fluctuation of arrival and service time. When the fluctuation is wilder, more resources are needed. We adopt EWMA to measure the value of $(\mu_a,\sigma_a)$ and $(\mu_s,\sigma_s)$. The long-timescale resource allocation algorithm at each LTI contains three steps.
	
	\emph{Step 1:} Find a set of frames for each flow, which contains all frames with the tolerable queuing time less than $\delta$.
	
	\emph{Step 2:} Based on C2, the algorithm uses dichotomy search to find the maximum target queuing delay for each flow.
	
	\emph{Step 3:} Based on Eq. (10), the algorithm calculates bandwidth resource allocation result for each flow.
	\begin{algorithm}[t]
		\caption{Short-timescale Resource Scheduling Algorithm }
		\begin{algorithmic}[1]
			\STATE Initialize $\mathcal F^0 = \emptyset, \mathcal F^1 = \emptyset,\left\{b_f^{(n,t)}\right\}_{f\in \mathcal F} = \left\{0\right\}$		
			\FOR{flow $f\in \mathcal F$}
			\IF {$V^{n,t}_{f}>V^{n,t-1}_f$}
			\STATE $\mathcal F^0=\mathcal F^0\cup \left\{f\right\}$
			\ELSE
			\STATE $\mathcal F^1=\mathcal F^1\cup\left\{f\right\}$
			\ENDIF
			\ENDFOR
			\WHILE {$\sum_{f}b_f^{(n,t)} \leq B-\sum_{f}\hat{b}_f^n$}
			\IF {$\mathcal F^0 \neq \emptyset$}
			\STATE Select a flow $f\in \mathcal F^0$
			\STATE $b_f^{n,t} = g(d_f^n- (V^{n,t}_{f}-V^{n,t-1}_f))-g(d_f^n)$
			\STATE $\mathcal F^0=\mathcal F^0 \backslash \left\{f\right\}$
			\ELSE
			\FOR{flow $f\in \mathcal F$}
			\STATE Find the frame $\mathsf{F}_i$ that satisfies $\mathsf{F}_i\in \mathcal {Q}_f^{(n,t)}\backslash\mathcal {A}_f^{(n,t)}$ and $\mathsf{F}_{i-1}\in \mathcal {A}_f^{(n,t)}$
			\STATE Calculate $\Delta U_f=U(b_f^{(n,t)}+\frac{s_i}{\delta/T})-U(b_f^{(n,t)})$
			\ENDFOR
			\STATE $f=\arg \max_f \Delta U_f$; $b_f^{(n,t)}=b_f^{(n,t)}+\frac{s_i}{\delta/T}$
			\ENDIF
			\ENDWHILE
			\RETURN $\left\{b_f^{(n,t)}\right\}_{f\in \mathcal F}$
		\end{algorithmic}
	\end{algorithm}
	\setlength{\textfloatsep}{7pt}
	\subsection{Short-Timescale Resource Scheduling}
	The goal of short-timescale resource scheduling is to cope with the network fluctuation which causes the true value of queuing delay bound varying. Thus, the queuing delay bound should be updated based on the RTT information carried by newly arrived frames. For each flow, the value of  $\widetilde{RTT}_{c^{'},m^{'},k^{'}}-\widetilde{q}_{c^{'},m^{'},k^{'}}$ among different frames can reflect the network fluctuation. Thus, we define the value from the frame of flow $f$ arriving at the beginning of STI $(n,t)$ as network state, denoted by $V^{n,t}_f$. The queuing delay bounds of frames $\mathsf{F}$ of flow $f$ should be revised to $DDL_{\mathsf{F}}-V^{n,t}_f$ based on Eq. (2). Besides, the available resource should be scheduled efficiently, i.e., flow with high resource utilization is expected to obtain more resources. Thus, we define $U(b_f^{n,t}) = \sum_{\mathsf{F}\in \mathcal {A}_f^{(n,t)}}\gamma_{\mathsf{F}}/(b_f^{n,t}+\hat{b}_f^n)$ as a utility function and optimize the utility subject to constraints in Eq. (7). The denominator is the total resources that flow $f$ can obtain in the STI $(n,t)$. The numerator is the sum of importance of frames that can be forwarded in the STI $(n,t)$, which is also related to the total allocated resources. Note that the utility function is non-convex. Besides, the relationship between utility function and resource scheduling decision is related to the importance of arriving frames, which varies in time and across flows. Thus, it is difficult to find an optimal solution to this problem. To reduce the complexity of solution, we propose a lightweight heuristic algorithm shown in Alg. 1. Firstly, although long-timescale resource allocation process is to assure the quality constraint, the variable queuing delay bound may cause the constraint unsatisfied, i.e., when the bound decreases. Secondly, the efficiency of resource scheduling should be considered. Thus, the algorithm contains two phases as follows.
	
	\emph{Phase 1 (resource compensation):} At each STI, we divide all flows into two sets based on the variation tend of queuing delay bound. If $V^{n,t}_{f}>V^{n,t-1}_f$, put flow $f$ into set $\mathcal F^0$, otherwise put flow $f$ into set $\mathcal F^1$ (line 2-8). As the decreasing queuing delay bounds of flows in set $\mathcal F^0$ leads to stricter target queuing delay, these flows are to be selected first and obtain more resources according to the relationship between resource allocation result $\hat{b}^n_f$ and maximum target queuing delay $d^n_f$ in Eq. (10), which is rewritten as $\hat{b}^n_f=g(d^n_f)$ for brevity. For each flow $f\in \mathcal F^0$, the maximum target queuing delay changes with the network state variation, i.e., $\Delta V^{n,t}_{f}=V^{n,t}_{f}-V^{n,t-1}_f$. Thus, the compensated resources for flow $f$ are $g(d_f^n- \Delta V^{n,t}_{f})-g(d_f^n)$ (line 11-13). 
	
	\emph{Phase 2 (efficiency optimization):} If there exist surplus resources after completing phase 1, a heuristic method is adopted to schedule them for maximizing the efficiency of resource utilization. At the beginning of efficiency optimization phase, we calculate utility function $U(b_f^{n,t})$ for each flow based on current total resource decision. Meanwhile, the gain of utility function in the case that one more frame is forwarded can be obtained (line 16-17), where $s_i$ means the data size of the frame and then $\frac{s_i}{\delta/T}$ represents the resource required to forward the frame. The flow whose gain is highest is selected and scheduled to the corresponding resource (line 19). Iterate the above procedure until resources are exhausted.
	
	\section{Performance Evaluation}
	In this section, we evaluate the performance of the proposed scheme. The following describes simulation setup and results.
	\subsection{Simulation Setup}
	We build a dumbbell network topology on the ns3 to simulate multiple VR video flows sharing a bottleneck link. The propagation delay of bottleneck link is 5ms and the data rate ranges from $25$ Mbps to $35$ Mbps. The number of flows is 10 and each of them is simulated based on a real VR video dataset\cite{xu2018gaze}. The length and frame rate of VR videos are $30$ s and $30$ fps. The duration of a chunk is $1$ s and each chunk is $4\times 6$ tiling. We adopt long short-term memory algorithm to predict viewport and obtain the viewing probability. The maximum packet size is 1500 bytes and the practical packet size depends on the frame size. For the parameters in our scheme, we set $\delta=1$s, $\delta/T =50$ms and $\beta = 0.01$.
	
	We compare our scheme with Round-Robin (RR) and Early Deadline First (EDF)\cite{huang2004qos} methods. RR schedules all flows equally. EDF selects the most urgent flow each time one frame is forwarded. In addtion, to test the performance of algorithms under varying network conditions, we generate data that follows exponential distribution to simulate the delay from the server to bottleneck node.
	\begin{figure}[t]
		\centering
		\setlength{\abovecaptionskip}{-0.05cm}
		\subfigcapskip = -2pt
		\subfigure[]{	
			
			\includegraphics[width=0.45\linewidth]{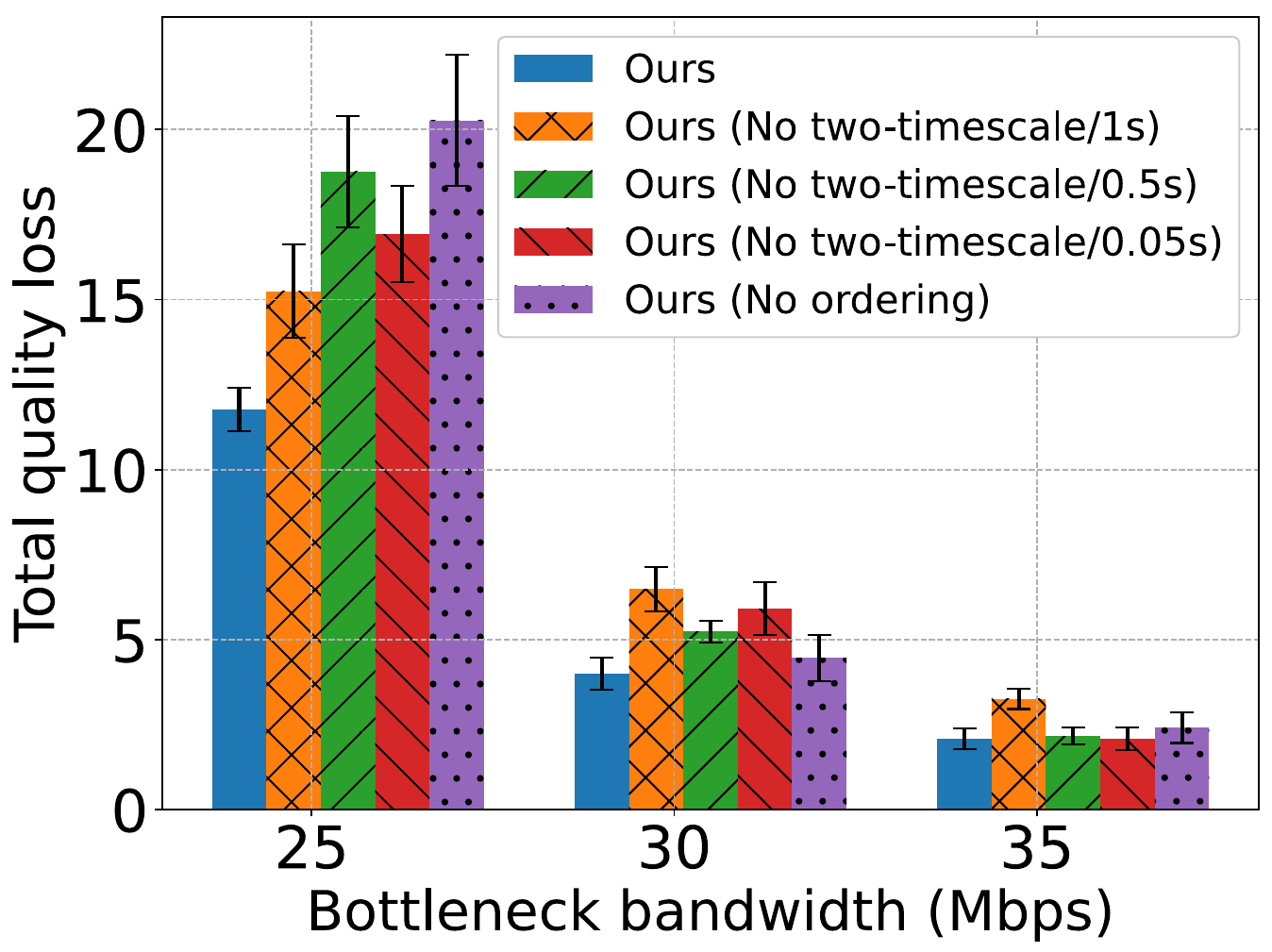}
		}
		\subfigure[]{	
			
			\includegraphics[width=0.45\linewidth]{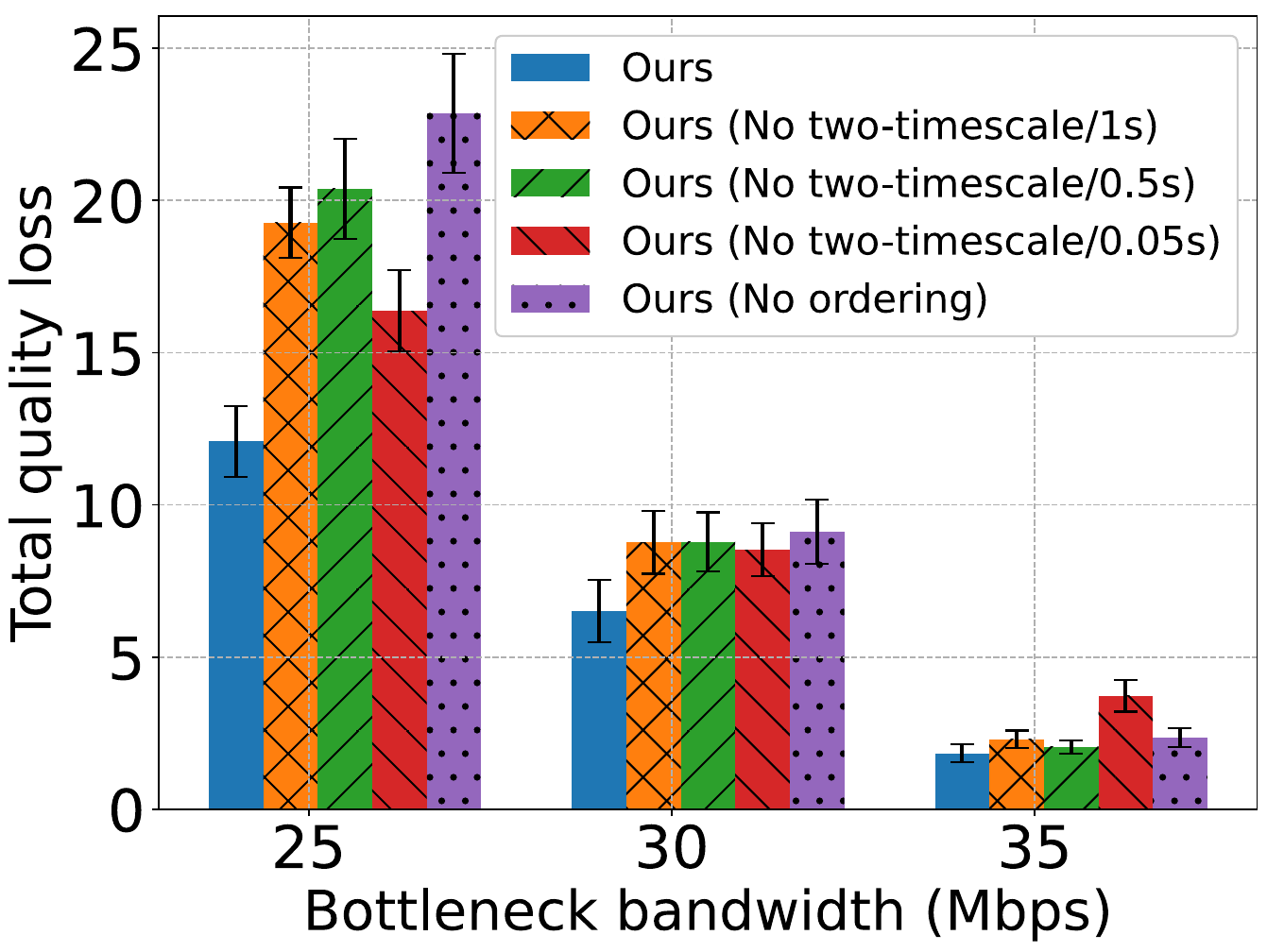}
		}
		\caption{Ablation study for the importance of two-timescale and frame ordering with different network state: $\left(a\right)$ stable and $\left(b\right)$ unstable. }
		\vspace{-0.12cm}
	\end{figure}
	\subsection{Ablation Study}
	To evaluate the effectiveness of different components in our scheme, we perform an ablation study by comparing with our scheme without frame ordering and two timescales. Fig. 4 shows the total quality loss (i.e., optimization objective in Eq. (7)) and standard deviation of all flows under different conditions, where No two-timescale/1s means that STI and LTI are both one second. To fully illustrate the necessity of two timescales, we test the performance of one-timescale scheme at $1$ s, $500$ ms and $50$ ms. We can see that removing any component will increase the quality loss and the loss gap among all flows. If removing frame ordering, the priority of different frames within each flow cannot be considered, leading to the performance degradation. As for running algorithms on single timescale, long interval cannot respond to the variation of traffic promptly. And short interval cannot assure the effect of Alg. 1 because the set $\mathcal {S}_f^{n}$ may be empty resulting in an inability to obtain actual resource demands. In addition, in most cases, as the congestion level deepens, the advantage of the overall scheme is more prominent.
	\begin{figure}[t]
		\centering
		\setlength{\abovecaptionskip}{-0.13cm}
		\subfigcapskip = -2pt
		\subfigure[]{	
			
			\includegraphics[width=0.45\linewidth]{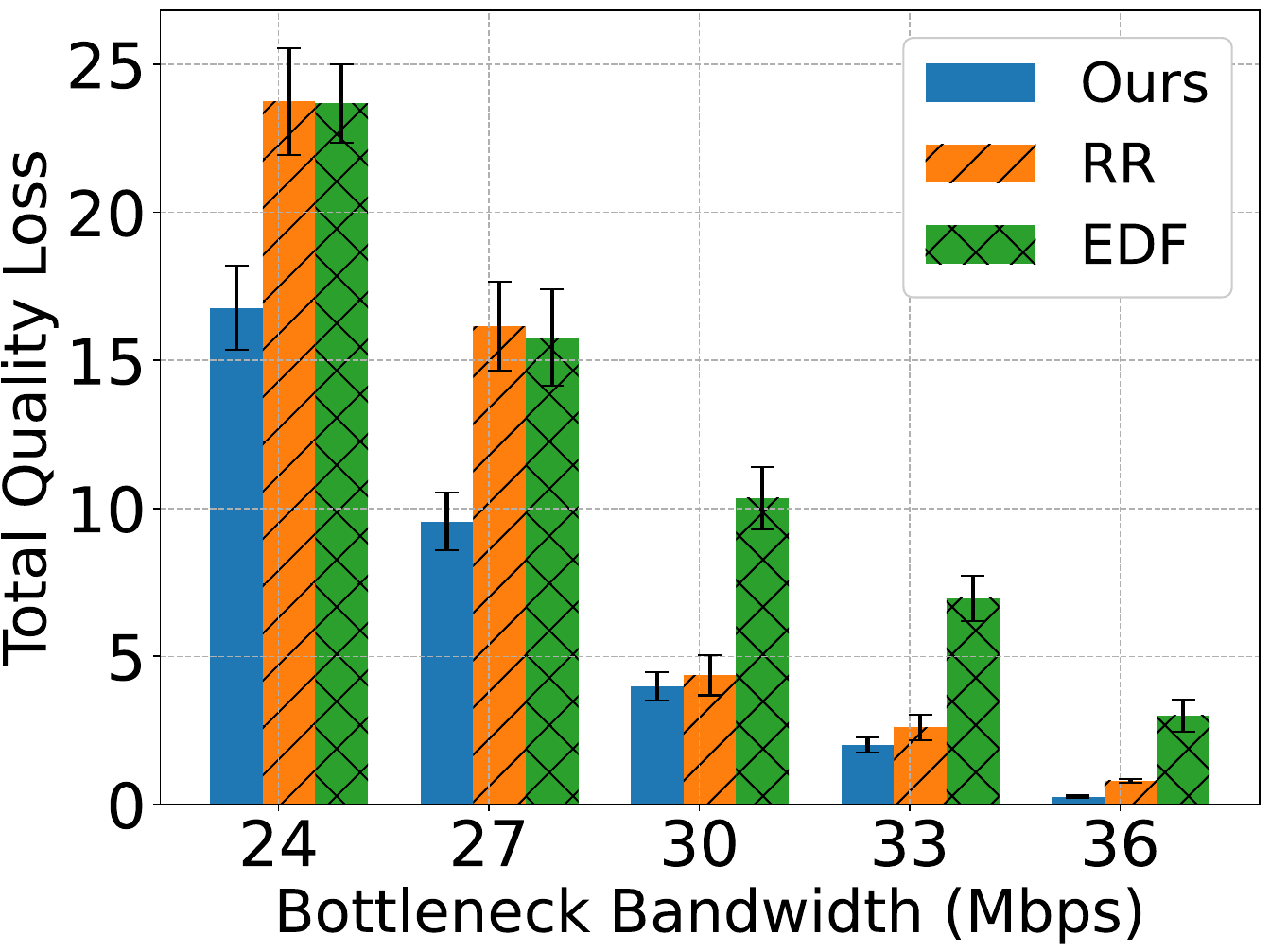}
		}
		\subfigure[]{	
			
			\includegraphics[width=0.45\linewidth]{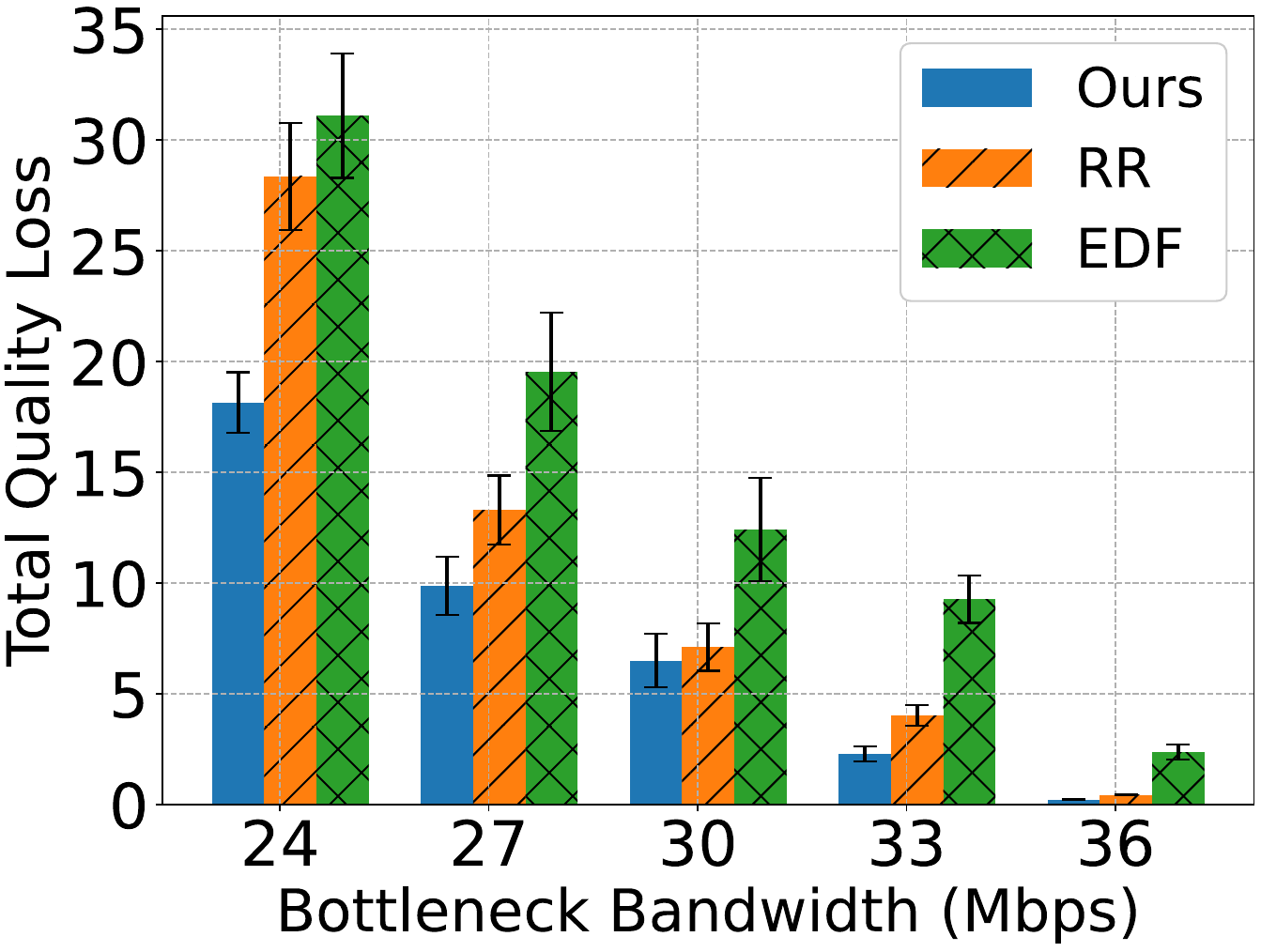}
		}
		\caption{Comparisons of total quality loss with different network state: $\left(a\right)$ stable and $\left(b\right)$ unstable. }
		\vspace{-0.3cm}
	\end{figure}
	\subsection{Performance Comparison}
	Figure 5 shows performance comparison results of total quality loss and standard deviation of all flows with different conditions. Our scheme can achieve lower quality loss and higher quality fairness in all cases. When network state is stable, our scheme can reduce at most $90\%$ overall quality loss and $94\%$ loss gap among different flows compared with the benchmarks. This is because we not only estimate the resource demands of each flow via queuing theory, but also consider frame importance. When network condition is unstable, our scheme can maintain a good performance through short-timescale adjustment and improve $12\%-72\%$ in quality loss. 
	
	In addition, we show the average frame drop rate of different methods in Fig. 6. Our scheme can achieve lower frame drop rate compared with the benchmarks. When network congestion is severe, our scheme can decrease $18\%-39\%$ drop rate. This indicates that resource allocation is more critical in this case, and our scheme can allocate proper resources for each flow.
	\begin{figure}[t]
		\centering
		\setlength{\abovecaptionskip}{-0.1cm}
		\subfigcapskip = -2pt
		\subfigure[]
		{
			\includegraphics[width=0.464\linewidth]{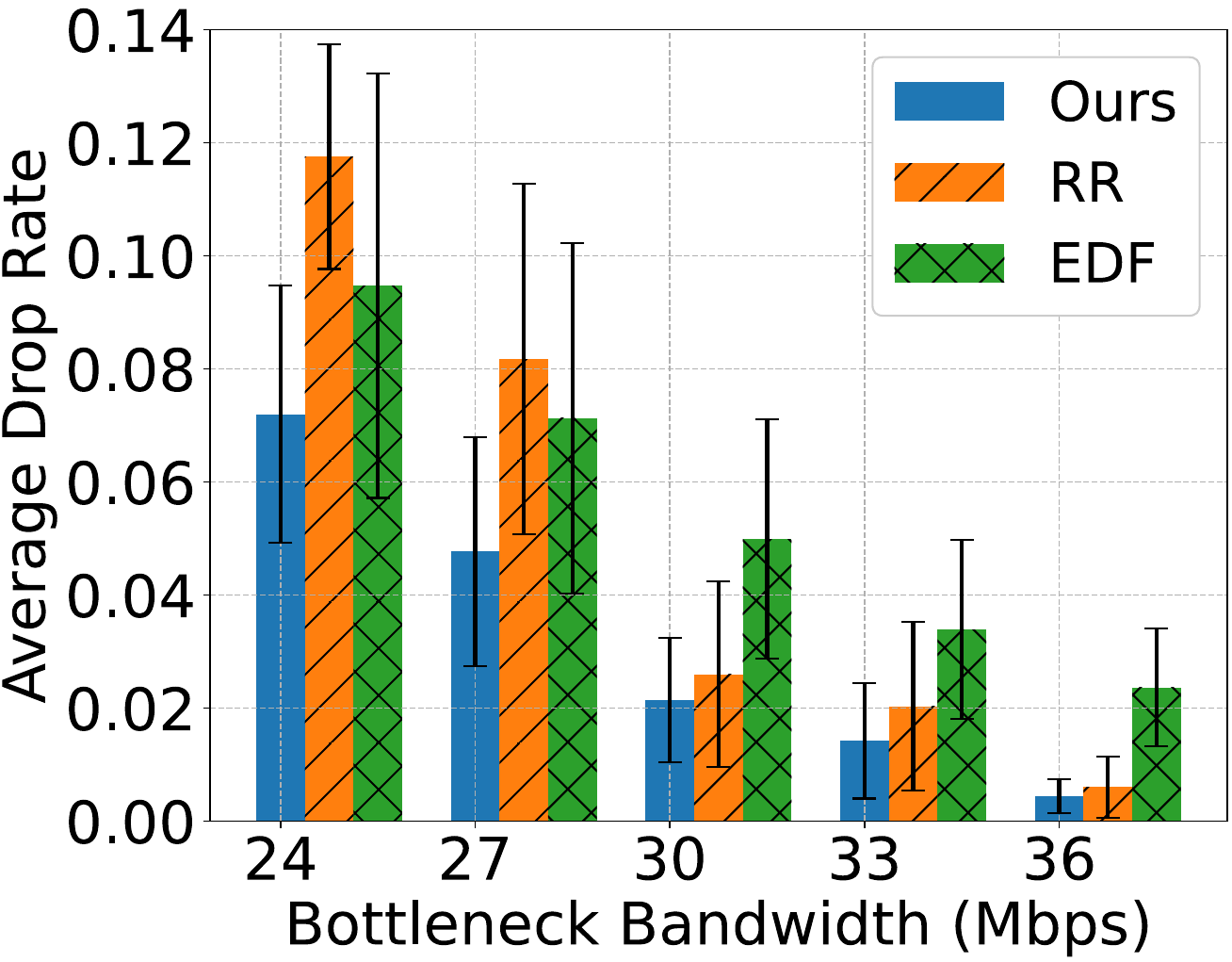}
		}
		\subfigure[]
		{
			\includegraphics[width=0.459\linewidth]{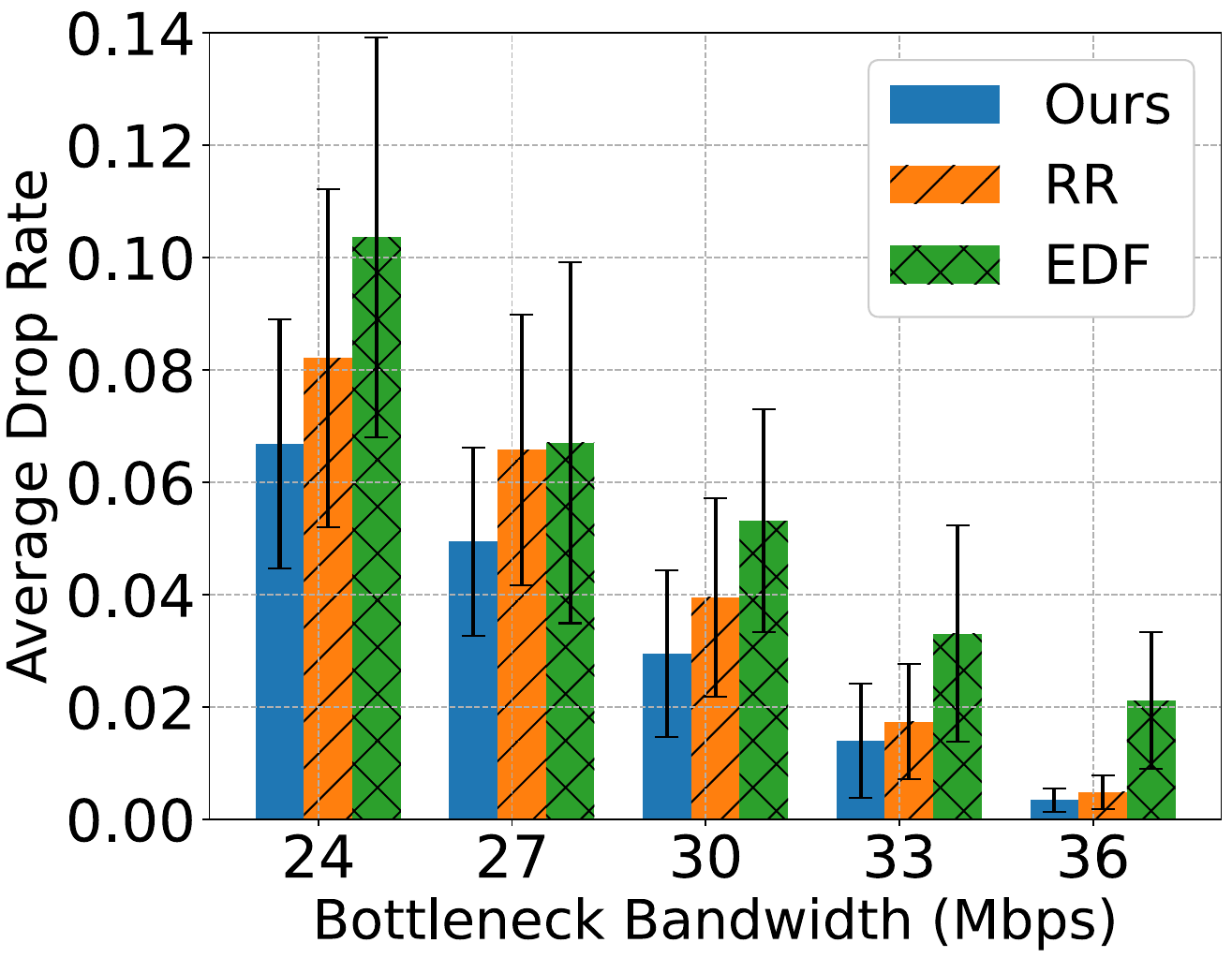}
		}
		\caption{Comparisons of average frame drop rate with different network state: $\left(a\right)$ stable and $\left(b\right)$ unstable.} 
		\label{12}
		\vspace{-0.1cm}
	\end{figure}
	\section{Conclusion}
	In this paper, we have proposed a deadline-aware two-timescale resource allocation scheme for multi-user VR video streaming in congestion scenario. First, we have designed a queuing delay bound estimation model to drop expired data proactively. Then, to measure the impact from the loss of different frames on QoE, we jointly considered viewport feature of VR video and encoding method to model the importance of frames. Based on frame queuing delay bound and importance, we adopted a frame sorting rule for each flow and proposed a two-timescale resource allocation scheme. On the long timscale, a queuing theory based resource allocation method was proposed to assure per-flow minimum quality requirement. On the short timescale, we fine-tuned the allocation results via a low-complexity heuristic algorithm. Simulation results have proved that the performance of the proposed scheme can significantly reduce quality loss and improve quality fairness compared with benchmark methods under different network conditions. For the future work, we will explore how to combine our scheme with adaptive transmission architecture.
	
	

\end{document}